\documentclass[12pt,twoside]{article}

\usepackage{times}
\usepackage{microtype}
\usepackage{amsfonts}
\usepackage{amsmath}
\usepackage{eucal}
\usepackage{amsthm}
\usepackage{amscd}
\usepackage{eucal}
\usepackage{cite}
\usepackage{graphicx}
\usepackage{a4}
\usepackage{booktabs}
\usepackage{array}

\def\beq{\begin{equation}}
\def\eeq{\end{equation}}
\def\bea{\begin{eqnarray}}
\def\eea{\end{eqnarray}}

\def\beann{\begin{eqnarray*}}
\def\eeann{\end{eqnarray*}}

\let\a=\alpha \let\be=\beta \let\g=\gamma 
\let\e=\varepsilon \let\z=\zeta \let\h=\eta \let\th=\theta

 \let\k=\kappa \let\la=\lambda \let\m=\mu
 \let\x=\xi \let\p=\pi \let\r=\rho \let\s=\sigma
\let\om=\omega 
\let\ph=\varphi   
\let\Om=\Omega  
 \let\G=\Gamma \let\D=\Delta

\let\qd=\quad \let\qqd=\qquad 

\def\epp{\, .}
\def\epc{\, ,}

\def\tst#1{{\textstyle #1}}

\theoremstyle{plain}

\newtheorem*{corollary*}{Corollary}

\newtheorem*{conjecture*}{Conjecture}

\theoremstyle{definition}

\renewcommand{\thefootnote}{\fnsymbol{footnote}}

\def\2{\frac{1}{2}} \def\4{\frac{1}{4}}

\def\6{\partial}

\def\+{\dagger}

\def\<{\langle} \def\>{\rangle}

\def\CH{{\cal H}} 
\def\CO{{\cal O}} 
 \def\CS{{\cal S}}

 \def\CW{{\cal W}}

\def\i{{\rm i}}

\def\rd{{\rm d}}
\def\re{{\rm e}}

\DeclareMathOperator{\sh}{sh}
\DeclareMathOperator{\ch}{ch}
\DeclareMathOperator{\tgh}{th}

\DeclareMathOperator{\sech}{sech}
\DeclareMathOperator{\cth}{cth}

\DeclareMathOperator{\tr}{tr}

\DeclareMathOperator{\End}{End}

\DeclareMathOperator{\spin}{{\mathbb S}}
\DeclareMathOperator{\spinflip}{{\mathbb J}}

\def\av{\mathbf{a}}
\def\bv{\mathbf{b}}
\def\cv{\mathbf{c}}
\def\bb{\mathbf{b}}
\def\bc{\mathbf{c}}

\def\fv{\mathbf{f}}
\def\hv{\mathbf{h}}

\def\kv{\mathbf{k}}

\def\fa{\mathfrak{a}}

\def\fb{\mathfrak{b}}
\def\fbq{\overline{\mathfrak{b}}}

\DeclareMathOperator{\trace}{Tr}

\DeclareMathOperator{\E}{e}

\renewcommand{\tilde}{\widetilde}

\allowdisplaybreaks

\begin{document}

\thispagestyle{empty}

\begin{center}

{\Large {\bf Short-distance thermal correlations in the XXZ chain\\}}

\vspace{8mm}

Herman E. Boos\footnote{e-mail: boos@physik.uni-wuppertal.de},
Jens Damerau\footnote[3]{e-mail: damerau@physik.uni-wuppertal.de},
Frank G\"{o}hmann\footnote[2]{e-mail:
goehmann@physik.uni-wuppertal.de},
Andreas Kl\"umper\footnote[4]{e-mail:
kluemper@uni-wuppertal.de}\\

\vspace{2mm}
Fachbereich C -- Physik, Bergische Universit\"at Wuppertal,
42097 Wuppertal, Germany\\

\vspace{6mm}
Junji Suzuki\footnote[5]{e-mail: sjsuzuk@ipc.shizuoka.ac.jp}\\
\vspace{2mm}
Department of Physics, Faculty of Science, Shizuoka University,
Ohya 836, Suruga, Shizuoka, Japan\\

\vspace{6mm}
Alexander Wei{\ss}e\footnote[6]{e-mail:
weisse@physik.uni-greifswald.de}\\
\vspace{2mm}
Institut f\"ur Physik, Universit\"at Greifswald, 17487 Greifswald,
Germany\\

\vspace{35mm}

{\large {\bf Abstract}}

\end{center}

\begin{list}{}{\addtolength{\rightmargin}{9mm}
               \addtolength{\topsep}{-5mm}}
\item
Recent studies have revealed much of the mathematical structure
of the static correlation functions of the XXZ chain. Here we
use the results of those studies in order to work out explicit
examples of short-distance correlation functions in the infinite chain.
We compute two-point functions ranging over 2, 3 and 4 lattice sites as
functions of the temperature and the magnetic field for various
anisotropies in the massless regime $- 1 < \D < 1$. It turns out
that the new formulae are numerically efficient and allow us to
obtain the correlations functions over the full parameter range
with arbitrary precision.
\\[2ex]
\end{list}

\clearpage

\section{Introduction}
In recent years considerable progress was achieved in the
understanding of the static correlation functions of local
operators in the XXZ spin chain. This concerns mostly the
ground state correlation functions at zero magnetic field
for which a hidden fermi\-onic structure (outside the free
Fermion point $\D = 0$) was identified in \cite{BJMST06b,BJMST07app,%
BJMST08app}. The exponential of an operator, bilinear in
two types of annihilation operators and acting on the space
of local operators on the spin chain, generates all local
correlation functions, while the space of local operators on
the chain is generated by the corresponding creation operators.
For the inhomogeneous chain this structure implies the
factorization of arbitrary correlation functions into sums
over products of neighbour correlators.

A generalization of the exponential form to finite temperatures
and finite longitudinal magnetic field was suggested in
\cite{BGKS07}. Without magnetic field the same Fermi operators
as in the ground state can still be used, and the modification of
the ground state result concerns only certain functions
multiplying the annihilation operators in the exponent.
For finite magnetic field, on the other hand, it seems that
the algebraic structure must be altered as well and a new Fermi-like
operator appears linearly in the exponent.

Here we shall work out the finite temperature formula of
\cite{BGKS07} in detail and compute the longitudinal and
transversal spin-spin correlation functions for up to four
lattice sites. We shall concentrate on the massless regime
$- 1 < \D < 1$, since the massive regime $\D > 1$
requires the calculation of different functions in the
analytical part of our work, which we postpone to a separate
publication.

\section{The density matrix of a chain segment}
We consider the XXZ chain with Hamiltonian
\begin{equation} \label{xxzham}
     \CH_N = J \sum_{j=-N+1}^N \Bigl( \s_{j-1}^x \s_j^x
             + \s_{j-1}^y \s_j^y + \D (\s_{j-1}^z \s_j^z - 1) \Bigr)
\end{equation}
in the antiferromagnetic regime ($J > 0$ and $\D > - 1$)\footnote{The
results of this and of the following section are valid for all $\D > - 1$.
Later on in sections \ref{sec:physpart} and \ref{sec:examples} we restrict
ourselves to the massless regime $- 1 < \D < 1$.}. Here the operators
$\s_j^\a$, $j = - N + 1, \dots, N$, $\a = x, y, z$, act locally as Pauli
matrices, $J$ is the exchange coupling, and $\D = \ch(\h) =
\2 (q + q^{-1})$ is the anisotropy parameter.

The XXZ Hamiltonian preserves the z-component of the total spin
\begin{equation} \label{totalsz}
     \CS_N^z = \tst{\2} \sum_{j=-N+1}^N \s_j^z \epp
\end{equation}
The equilibrium properties of the chain at temperature $T$ and
external magnetic field $h$ are determined by the statistical
operator
\begin{equation} \label{statop}
     \r_N (T,h) = \frac{\re^{- (\CH_N - h \CS_N^z)/T}}
                       {\tr_{-N+1, \dots, N} 
		        \re^{- (\CH_N - h \CS_N^z)/T}} \epp
\end{equation}

We are interested in the system in the thermodynamic limit
$L = 2N \rightarrow \infty$. To perform this limit in a sensible
way we fix a positive integer $n$ and define
\begin{equation} \label{densmatgen}
     D_n (T,h) = \lim_{N \rightarrow \infty}
                 \tr_{-N+1, \dots, -1, 0, n+1, n+2, \dots, N} \,
                 \r_N (T,h) \epc
\end{equation}
the density matrix of the segment $[1,n]$ of the infinite chain.
By construction it satisfies the reduction property
\begin{equation} \label{densredu}
     \tr_n D_n (T,h) = D_{n - 1} (T,h) \epp
\end{equation}
Denote by $\CW$ the space of operators with non-trivial action
only on a finite number of positive lattice sites. Because of the
translational invariance of the Hamiltonian it is sufficient to 
consider this space. For $\CO \in \CW$ we write its restriction
to the first $n$ lattice sites as $\CO_{[1,n]}$. Then
\begin{equation} \label{unidens}
     D^*_{T,h} (\CO) = \lim_{n \rightarrow \infty}
         \tr_{1, \dots, n} \bigl( D_n (T,h) \CO_{[1,n]} \bigr)
\end{equation}
is well defined because of (\ref{densredu}) and determines the
thermal average of the operator $\CO$ in a magnetic field $h$,
which we shall also denote $\<\CO\>_{T, h}$. Mathematically
(\ref{unidens}) defines $D^*_{T,h}$ as a functional on $\CW$.
Along with $D^*_{T,h}$ we often study its inhomogeous generalization
(see \cite{BGKS07}) which we denote by the same letter. When acting
on an operator of length $n$ it depends on $n$ so-called inhomogenieties
$\x_j$, $j = 1, \dots, n$.

In our previous work \cite{BGKS07} we conjectured a rather
explicit formula for $D^*_{T,h}$ which was inspired in part
by \cite{BJMST06b}. The conjecture says that
\begin{equation} \label{main2}
     D^\ast_{T,h} (\CO)
        = \mathbf{tr} \bigl(\re^{\Om_1 + \Om_2} (\CO)\bigr) \epc
\end{equation}
where $\mathbf{tr}$ is the trace functional defined by
\begin{equation} \label{tracefun}
     \mathbf{tr}(\CO) = \dots \tst{\2} \tr_1\
                        \tst{\2} \tr_2\ \tst{\2} \tr_3 \dots (\CO)
                        \epc
\end{equation}
and the $\Om_j: \CW \rightarrow \CW$, $j = 1, 2$ are linear operators
of the form\footnote{We have slightly changed our notation for operators.
Instead of those of \cite{BGKS07} we rather employ the conventions of
the more recent paper \cite{BJMST08app}.}
\begin{subequations}
\label{Omega1and2}
\begin{align} \label{Omega1}
     \Om_1 & = - \lim_{\a \rightarrow 0}
                 \int_\G \int_\G \frac{\rd \z_1^2}{2 \p \i \z_1^2}
                 \frac{\rd \z_2^2}{2 \p \i \z_2^2} \:
		 \om(\m_1,\m_2;\a) \bb (\z _1,\a) \bc (\z _2,\a-1)
		 \epc \\[1ex]
     \Om_2 & = - \lim_{\a \rightarrow 0}
                 \int_\G \frac{\rd \z_1^2}{2 \p \i}
                 \: \ph(\m_1;\a) \mathbf{h} (\z_1, \a)
		 \epp \label{Omega2}
\end{align}
\end{subequations}
Here $\m_j = \ln \z_j$, $j = 1, 2$. As a consequence of taking the
limit $\a \rightarrow 0$ we shall only need the values of the function
$\om(\m_1, \m_2; \a)$ and of its $\a$-derivative at $\alpha = 0$.
Similarly we only need $\ph (\m_1; 0)$. All the functions that are
necessary for the actual evaluation of (\ref{Omega1and2})
will be introduced in section \ref{sec:physpart} below. They depend
on the temperature and on the magnetic field and describe what we call
the physical part of the problem. By way of contrast, the operators
$\bb (\z, \a)$, $\bc (\z, \a)$ and $\mathbf{h} (\z, \a)$ are
independent of $T$ and $h$. These operators are rational functions of
$\z^2$ and constitute the algebraic part of the problem. We will
explain their construction in the next section. The closed contour
$\G$ encircles the poles of the operators at the inhomogeneities
$\x_j^2$ (see below) and excludes all other singularities of the
integrand.

\section{The algebraic part of the construction} \label{sec:algpart}
We should first explain that our present status of understanding is
different for $\Om_1$ and $\Om_2$ in (\ref{Omega1and2}). The algebraic
part of $\Om_1$, built up from the operators $\bb (\z, \a)$ and
$\bc (\z, \a)$, is the same as for the ground state at vanishing
magnetic field. First versions of these operators were introduced
in \cite{BJMST06b}. They were optimized and further studied in
\cite{BJMST07app,BJMST08app}. Their analytic and algebraic structure
is by now well understood. In particular, they satisfy a reduction
property and anticommutation relations. The operator $\mathbf{h}
(\z, \a)$ was introduced in \cite{BGKS07} in order to account for
correlation functions odd under spin reversal. This operator is
less well studied. We plan to further elaborate on its mathematical
structure elsewhere. Below, when we work out examples of correlation
functions, we use some of its properties for $\a \rightarrow 0$ that
we verified by explicit calculations for up to four lattice sites.

In this section we follow to a certain extent the recent paper
\cite{BJMST08app} to which we refer the reader for further details.
For the definition of the operators $\bb (\z, \a)$, $\bc (\z,\a)$
we first of all generalize the space of local operators. Instead of
local operators we consider operators of the form
\begin{equation}
     q^{2 \a S (0)} {\cal O} \epc \qqd
        S(0) = \tst{\2} \sum_{j = - \infty}^0 \s_j^z \epc
\end{equation}
where $\cal O$ is local and $\a \in {\mathbb C}$. These operators
are called quasi-local with tail $\a$. They span a vector space
${\cal W}_\a$. The parameter $\a$ is introduced in order to regularize
the various objects introduced below. Some of them are rather singular
for $\a \rightarrow 0$, and the limit is, in general, only well
defined for appropriate combinations.

In \cite{BJMST08app} the operators $\bb(\z, \a)$, $\bc(\z, \a)$ 
were defined in two steps. In step one endomorphisms $\bb_{[kl]}
(\z, \a)$ and $\bc_{[kl]} (\z, \a)$ acting on $M_{[k,l]} =
\End (V_k \otimes \dots \otimes V_l)$ were defined, where every
$V_j \cong {\mathbb C}^2$ is the space of states of a spin in the
infinite chain. In step two a certain reduction property was used
to extend their action inductively to $\CW_\a$. Here we shall
only review step one, since we will be dealing with small finite
chain segments anyway, and we refer the reader to \cite{BJMST08app}
for the reduction property.

The operators $\bb_{[kl]} (\z, \a)$ and $\bc_{[kl]} (\z, \a)$ are
constructed from weighted traces of the elements of certain monodromy
matrices related to $U_q (\widehat{\mathfrak{sl}_2})$. These
monodromy matrices are products of two types of $L$-matrices with
two-dimensional and with infinite dimensional auxiliary space,
respectively.

The $L$-matrices with two-dimensional auxiliary space are directly
related to the $R$-matrix of the six-vertex model,
\begin{align}
     & R(\zeta) = \begin{pmatrix}
		     1&0&0&0\\
		     0&\be(\z)&\g(\z)&0\\
		     0&\g(\z)&\be(\z)&0\\
		     0&0&0&1
		  \end{pmatrix} \epc
\end{align}
where
\begin{equation}
     \be(\z) = \frac{(1 - \z^2)q}{1 - q^2 \z^2} \epc \qd
     \g(\z) = \frac{(1 - q^2)\z}{1 - q^2 \z^2} \epp
\end{equation}
Fixing an auxiliary space $V_a$ isomorphic to ${\mathbb C}^2$ we
define $L_{a,j} (\z) = \r (\z) R_{a,j} (\z)$, where $\r (\z)$ is
a scalar factor to be specified below. This is the standard $L$-matrix
of the six-vertex model. The corresponding monodromy matrix is
\begin{equation}
     T_{a,[k,l]} (\z) = L_{a,l} (\z/\x_l) \dots L_{a,k} (\z/\x_k) \epp
\end{equation}
It acts on $V_a \otimes V_k \otimes \dots \otimes V_l$. We
are interested in operators acting on $M_{[k,l]}$ or on $M_a \otimes
M_{[k,l]}$, where $M_a = \End(V_a)$. Such type of operators are naturally
given by the adjoint action of operators acting on $V_k \otimes \dots
\otimes V_l$ or on $V_a \otimes V_k \otimes \dots \otimes V_l$. An
example is the $L$-operator ${\mathbb L}_{a,j} (\z)$ defined by
\begin{equation}
     {\mathbb L}_{a,j} (\z)\, X_{[k,l]}
        = L_{a,j} (\z) X_{[k,l]} L_{a,j}^{-1} (\z)
\end{equation}
for $j \in \{k, \dots, l\}$ and for all $X_{[k,l]} \in M_a \otimes
M_{[k,l]}$. These $L$-operators generate the first type of monodromy
matrices entering the constructions of $\bb(\z, \a)$, $\bc(\z, \a)$,
\begin{equation} \label{monofund}
     {\mathbb T}_a (\z, \a) X_{[k,l]} =
        {\mathbb L}_{a,l} (\z/\x_l) \dots {\mathbb L}_{a,k} (\z/\xi_k)
	q^{\a \s_a^z} X_{[k,l]}
\end{equation}
for all $X_{[k,l]} \in M_a \otimes M_{[k,l]}$.

A second type of monodromy matrix acts adjointly on an infinite
dimensional bosonic auxiliary space, more precisely on a representation
space of the $q$-oscillator algebra $Osc$ defined in terms of generators
$\av$, $\av^*$, $q^{\pm D}$ and relations
\begin{equation}
     q^D \av q^{-D} = q^{-1} \av \epc \qd
     q^D \av^* q^{-D} = q \, \av^* \epc \qd
     \av \av^* = 1 - q^{2D + 2} \epc \qd
     \av^* \av = 1 - q^{2D} \epp
\end{equation}
Let $Osc_A$ a copy of $Osc$. We define
\begin{equation} \label{Losc}
     L_{A,j} (\z) = \s (\z)
                    \begin{pmatrix}
		    1 - \z^2 q^{2D_A + 2} & - \z \av_A \\
		    - \z \av^*_A & 1
                    \end{pmatrix}_j
                    \begin{pmatrix}
		    q^{- D_A} & 0 \\ 0 & q^{D_A}
                    \end{pmatrix}_j
\end{equation}
which acts on $Osc_A \otimes V_k \otimes \dots \otimes V_l$. The scalar
factor $\s(\z)$ is a convenient normalization and is required to solve
the functional equation
\begin{equation}
     \s (\z) \s (q^{-1} \z) = \frac{1}{1 - \z^2} \epp
\end{equation}
It also fixes the scalar factor in the definition of $L_{a,j} (\z)$,
\begin{equation}
     \r (\z) = \frac{q^{-1/2} \s(q^{-1} \z)}{\s(\z)} \epp
\end{equation}
For the $L$-operator (\ref{Losc}) we define again its adjoint action
\begin{equation}
     {\mathbb L}_{A,j} (\z)\, X_{[k,l]}
        = L_{A,j} (\z) X_{[k,l]} L_{A,j}^{-1} (\z) \epc
\end{equation}
for $j \in \{k, \dots, l\}$ and for all $X_{[k,l]} \in \End(Osc_A)
\otimes M_{[k,l]}$, and also the corresponding monodromy matrix
\begin{equation} \label{monoosc}
     {\mathbb T}_A (\z, \a) X_{[k,l]} =
        {\mathbb L}_{A,l} (\z/\x_l) \dots {\mathbb L}_{A,k} (\z/\xi_k)
	q^{2 \a D_A} X_{[k,l]} \epp
\end{equation}

The monodromy matrices (\ref{monofund}) and (\ref{monoosc}) are the
basic ingredients in the definition of the fermionic operators
entering the exponential form of the density matrix. In addition to
the monodromy matrices we shall need certain operators acting on
the spin. We introduce the adjoint action of the $z$-component of the
total spin on $M_{[k,l]}$,
\begin{equation}
     {\mathbb S} \, X_{[k,l]} = [S^z_{[k,l]},X^{}_{[k,l]}] \epc \qd
        S^z_{[k,l]} = \tst{\2} \sum_{j = k}^l \s^z_j \epc
\end{equation}
and the spin reversal operator defined by
\begin{equation}
     {\mathbb J} \, X_{[k,l]} = \Bigl[ \prod_{j = k}^l \s^x_j \Bigr]
        X_{[k,l]} \Bigl[ \prod_{j = k}^l \s^x_j \Bigr] \epp
\end{equation}
Somewhat loosely we say that an operator $X_{[k,l]}$ has spin $s$ if
\begin{equation}
     {\mathbb S} \, X_{[k,l]} = s X_{[k,l]} \epp
\end{equation}
We call $X_{[k,l]}$ spin-reversal even if ${\mathbb J} \, X_{[k,l]}
= X_{[k,l]}$ and spin-reversal odd if ${\mathbb J} \, X_{[k,l]} =
- X_{[k,l]}$.

The annihilation operators $\bb(\z, \a)$ and $\bc(\z, \a)$ as well as
a couple of other interesting fermionic operators (see \cite{BJMST08app})
can be derived from a master object $\kv (\z, \a)$ defined as
\begin{equation}
     \kv (\z, \a) \, X_{[k,l]}
        = \tr_{A, a} \Bigl\{ \s_a^+ {\mathbb T}_a (\z, \a)
	             {\mathbb T}_A (\z, \a) \z^{\a - {\mathbb S}}
		     \bigl(q^{- 2S^z_{[k,l]}} X_{[k,l]}\bigr) \Bigr\}
                     \epp
\end{equation}
Here the bosonic trace is understood in such a way that it agrees
with the trace in the bosonic representation $W^+ = \oplus_{k \ge 0}
{\mathbb C} |k\>$ for $|q| \le 1$ but not a root of unity (with
$q^D |k\> = q^k |k\>, \av |k\> = (1 - q^{2k}) |k - 1\>, \av^* |k\> =
|k + 1\>$). For more details see appendix A of \cite{BJMST08app}.

We shall give a description of the operators $\bb(\z, \a)$ and
$\bc(\z, \a)$ in terms of certain residua of $\kv (\z, \a)$ which
is appropriate for implementing them on a computer. Let us fix an
operator $X_{[k,l]}$ of spin $s$. Note that $|s| \le l - k + 1 = n$.
The following statements about the analytic structure of
$\kv (\z, \a)$ were worked out in \cite{BJMST08app}.
\begin{enumerate}
\item
$\z^{- \a + s - 1} \kv (\z, \a) X_{[k,l]}$ is rational in $\z^2$,
\item
Its only singularities in the finite complex plane are poles at
$\z^2 = \x_j^2, q^{\pm 2} \x_j^2$,
\item
$\z^{- \a - s + 1} \kv (\z, \a) X_{[k,l]}$ is regular at $\z^2 =
\infty$.
\end{enumerate}
It follows that $\kv_{\rm skal} (\z, \a) X_{[k,l]} = \z^{- \a - s - 1}
\kv (\z, \a) X_{[k,l]}$ is regular at $\z^2 = 0$ if $s \le 0$  and
has at most a pole of order $s$ in $\z^2$ at $\z^2 = 0$ if $s > 0$.
Furthermore $\kv_{\rm skal} (\z, \a) X_{[k,l]} \sim \z^{-2}$ for
$\z \rightarrow \infty$. Hence,
\begin{equation} \label{kskal}
     \kv_{\rm skal} (\z, \a) X_{[k,l]}
        = \biggl[ \sum_{j=1}^n \sum_{\e = 0, \pm}
	  \frac{\r_j^{(\e)} (\a)}{\z^2 - q^{2 \e} \x_j^2}
        + \sum_{j=1}^s \z^{-2j} \, \k_j (\a) \biggr] X_{[k,l]}
\end{equation}
(where the second sum over $j$ is zero by definition for $s < 1$).
The residua $\r_j^{(\e)} (\a)$ and the Laurent coefficients $\k_j (\a)$
determine all those operators that are needed for vanishing magnetic
field. They are \emph{defined} by equation (\ref{kskal}).

First of all there are the operators $\bar \cv(\z,\a)$, $\cv(\z,\a)$
and $\fv(\z,\a)$ defined in section 2.7 of \cite{BJMST08app}. In
terms of the residua and Laurent coefficients they read
\begin{subequations}
\begin{align}
     \bar \cv(\z,\a) X_{[k,l]} & =
        \z^{\a + s + 1} \sum_{j=1}^n \frac{\z^2 + \x_j^2}{\z^2 - \x_j^2}
        \cdot \frac{\r_j^{(0)} (\a) X_{[k,l]}}{2 \x_j^2} \epc \\[1ex]
     \cv(\z,\a) X_{[k,l]} & =
        \z^{\a + s + 1} \sum_{j=1}^n \sum_{\e = \pm}
	\frac{\z^2 + \x_j^2}{\z^2 - \x_j^2}
        \cdot \frac{q^{\e (\a + s - 1)} \r_j^{(\e)} (\a) X_{[k,l]}}
	           {4 \x_j^2} \epc \\[1ex]
     \fv(\z,\a) X_{[k,l]} & = \z^{\a + s + 1} \biggl[
        \sum_{j=0}^s \frac{\z^{-2j} \k_j (\a) X_{[k,l]}}
	                  {q^{\a + s + 1 - 2j} - q^{- \a - s - 1 + 2j}}
			  \notag \\[-1ex] & \mspace{136.mu}
      - \sum_{j=1}^n \sum_{\e = \pm}
	\frac{\z^2 + \x_j^2}{\z^2 - \x_j^2}
        \cdot \frac{\e \, q^{\e (\a + s - 1)} \r_j^{(\e)} (\a) X_{[k,l]}}
	           {4 \x_j^2} \biggr] \epc
\end{align}
\end{subequations}
where by definition
\begin{equation}
     \k_0 = - \sum_{j=1}^n \sum_{\e = 0, \pm}
              \frac{\r_j^{(\e)} (\a)}{2 q^{2 \e} \x_j^2} \epp
\end{equation}
In terms of these operators $\kv (\z, \a)$ is decomposed as
\begin{equation}
     \kv (\z, \a) X_{[k,l]} =
        \bigl( \bar \cv(\z,\a) + \cv(q \z,\a) + \cv(q^{-1} \z,\a)
               + \fv(q \z,\a) - \fv(q^{-1} \z,\a) \bigr) X_{[k,l]} \epp
\end{equation}

The operator $\bv(\z,\a)$ that is needed in the construction of
the density matrix is defined as
\begin{equation}
     \bv (\z,\a) = q^{-1}(q^{\a - \spin - 1} - q^{- \a + \spin + 1})
                   \spinflip \cv (\z,- \a) \spinflip \epc
\end{equation}
where $\spinflip$ is the spin-flip in the adjoint representation.
Taking into account that $\spinflip \spin \spinflip = - \spin$
we find
\begin{equation}
     \bv(\z,\a) X_{[k,l]} =
        \z^{- \a - s + 1} \sum_{j=1}^n \sum_{\e = \pm}
	\frac{\z^2 + \x_j^2}{\z^2 - \x_j^2} \cdot
	\frac{q^{- \e (\a + s + 1)} \tilde \r_j^{(\e)} (\a) X_{[k,l]}}
             {4 \x_j^2} \epc
\end{equation}
where
\begin{equation}
     \tilde \r_j^{(\e)} (\a) X_{[k,l]} = q^{-1}(q^{\a - s} - q^{s - \a})
        \spinflip \r_j^{(\e)} (- \a) \spinflip X_{[k,l]} \epc
\end{equation}
for $\e = \pm$.

For later convenience we introduce the operators
\begin{equation}
     \bv_j (\a) = \sum_{\e = \pm}
        \frac{q^{- \e (\a + \spin + 2)} \tilde \r_j^{(\e)} (\a)}
	     {2 \x_j^2} \epc \qd
     \cv_j (\a) = \sum_{\e = \pm}
        \frac{q^{\e (\a + \spin - 2)} \r_j^{(\e)} (\a)}
	     {2 \x_j^2} \epc
\end{equation}
for $j = 1, \dots, n$. In terms of these operators we have
\begin{subequations}
\begin{align}
     \bv(\z,\a) X_{[k,l]} & =
        \z^{- \a - s + 1} \sum_{j=1}^n
	\2 \frac{\z^2 + \x_j^2}{\z^2 - \x_j^2}
        \cdot \bv_j (\a) X_{[k,l]} \epc \\[1ex]
     \cv(\z,\a) X_{[k,l]} & =
        \z^{\a + s + 1} \sum_{j=1}^n
	\2 \frac{\z^2 + \x_j^2}{\z^2 - \x_j^2}
        \cdot \cv_j (\a) X_{[k,l]} \epp
\end{align}
\end{subequations}

Let $\omega (\m_1, \m_2; \a)$ denote the function that determines
the physical part of the problem. $\omega (\m_1, \m_2; \a)$ is a
known function of the temperature and the magnetic field (see section 3
of \cite{BGKS07}). Here we only need its property
$\omega (\m_2, \m_1; \a) = \omega (\m_1, \m_2; - \a)$. We also set
$\tilde \omega (\m_1, \m_2; \a) = \omega (\m_1, \m_2; \a) (\z_2/\z_1)^\a$
(recall that $\z_j = \re^{\m_j}$) which has the same symmetry. Then,
by definition,
\begin{multline} \label{omal}
     \Om_1 (\a) X_{[k,l]} =
        - \int_\G \frac{\rd \z_1^2}{2 \p \i \z_1^2}
	  \int_\G \frac{\rd \z_2^2}{2 \p \i \z_2^2}
	  \omega (\m_1, \m_2; \a) \bv (\z_1, \a) \cv (\z_2, \a - 1)
	  X_{[k,l]} \\[1ex] =
        - \sum_{1 \le i < j \le n}
          \bigl[ \tilde \omega (\la_i, \la_j; \a)
	     \bv_i (\a) \cv_j (\a - 1) + \tilde \omega (\la_j, \la_i; \a)
	     \bv_j (\a) \cv_i (\a - 1) \bigr] X_{[k,l]}
\end{multline}
(where $\la_j = \ln \x_j$) for all $X_{[k,l]}$ with $\spin(X_{[k,l]})
= 0$. The closed contour $\G$ in (\ref{omal}) encircles all the
$\x_j^2$ (and excludes the singularities of $\om$). In the second
equation we used that
\begin{equation}
     \bv_i (\a) \cv_i (\a - 1) = 0 \epc
        \qd \text{for $i = 1, \dots, n$} \epp
\end{equation}

For the physical correlation functions the limit $\a \rightarrow 0$
remains to be done. We shall write $\Om_1 = \Om_1 (0)$. We claim that
\begin{equation} \label{om1resform}
     \Om_1 X_{[k,l]} = - \sum_{1 \le i < j \le n} \bigl[
                       \omega (\la_i, \la_j) \Om_{i,j}^+ +
                       \omega' (\la_i, \la_j) \Om_{i,j}^-
		       \bigr] X_{[k,l]} \epc
\end{equation}
where
\begin{subequations}
\label{ompmij}
\begin{align}
     \Om_{i,j}^+ = & \lim_{\a \rightarrow 0}
                       \bigl[ \bv_i (\a) \cv_j (\a - 1)
		            + \bv_j (\a) \cv_i (\a - 1) \bigr]
                     \epc \\[1ex]
     \Om_{i,j}^- = & \lim_{\a \rightarrow 0} \, \a
                       \bigl[ \bv_i (\a) \cv_j (\a - 1)
		            - \bv_j (\a) \cv_i (\a - 1) \bigr] \epc
\end{align}
\end{subequations}
and
\begin{equation}
     \omega (\la_i, \la_j)
        = \tilde \omega (\la_i, \la_j; 0) \epc \qd
     \omega' (\la_i, \la_j)
        = \6_\a \tilde \omega (\la_i, \la_j; \a) \bigr|_{\a = 0} \epp
\end{equation}
The existence of the limits in (\ref{ompmij}) was proved in
\cite{BGKS07}.

As was already mentioned in the introduction to this section, the
operator $\hv (\z, \a)$ introduced in \cite{BGKS07} is a less
well understood than $\bv (\z, \a)$ and $\cv (\z, \a)$. Here we
adapt the definition of \cite{BGKS07} to the novel conventions of
\cite{BJMST08app} and set
\begin{equation}
     \hv (\z, \a) \, X_{[k,l]}
        = (1 - q^\a) \tr_{A, a} \Bigl\{ \s_a^+ {\mathbb T}_a (\z, \a)
	             {\mathbb T}_A (\z, \a) \z^{\a - {\mathbb S}}
		     \bigl(q^{- 2S^z_{[k,l]}} \av \,
                     X_{[k,l]}\bigr) \Bigr\} \epp
\end{equation}
This operator has some similarity with the operator $\kv (\z, \a)$
introduced above. The only differences are the prefactor $(1 - q^\a)$
and the insertion of an extra bosonic  annihilation operator on the right
hand side. For this reason its analytic structure as a function of
$\z$ is rather similar to that of $\kv (\z, \a)$. Following the
same reasoning as in section 2.5 of \cite{BJMST08app} and using that
$[\spin, \hv (\z, \a)] = 0$ one sees that for an operator $X_{[k,l]}$
of spin $s$
\begin{enumerate}
\item
$\z^{- \a + s} \, \hv (\z, \a) X_{[k,l]}$ is rational in $\z^2$,
\item
its only singularities in the finite complex plane are poles at
$\z^2 = \x_j^2, q^{\pm 2} \x_j^2$,
\item
$\z^{- \a - s + 2} \, \hv (\z, \a) X_{[k,l]}$ is regular at $\z^2 =
\infty$.
\end{enumerate}
It follows that $\hv (\z, \a) X_{[k,l]}$ is of the form
\begin{equation} \label{partialh}
     \hv (\z, \a) X_{[k,l]}
        = \z^{\a + s} \Biggl[ \sum_{j=1}^n
	  \biggl\{ \frac{\hv_j (\a)}{\z^2 - \x_j^2} +
          \sum_{\e = \pm}
	  \frac{\hv_j^{(\e)} (\a)}{\z^2 - q^{2 \e} \x_j^2} \biggr\}
        + \sum_{j=1}^s \z^{-2j} \, \th_j (\a) \Biggr] X_{[k,l]} \epp
\end{equation}

We denote the function that determines the physical part of $\Om_2$
by $\ph (\m; \a)$ (see \cite{BGKS07}) and set $\tilde \ph (\m; \a)
= \ph(\m; \a) \z^\a$. Then, for every $X_{[k,l]}$ of spin 0,
\begin{multline} \label{om2resform}
     \Om_2 X_{[k,l]} = - \lim_{\a \rightarrow 0}
               \int_\G \frac{\rd \z_1^2}{2 \p \i}
               \: \ph(\m_1;\a) \mathbf{h} (\z_1, \a) X_{[k,l]} = \\
               - \lim_{\a \rightarrow 0} \sum_{j = 1}^n
                 \tilde \ph (\la_j; \a) \hv_j (\a) X_{[k,l]} =
               - \sum_{j = 1}^n \ph (\la_j) \hv_j X_{[k,l]} \epc
\end{multline}
where
\begin{equation}
     \hv_j = \lim_{\a \rightarrow 0} \hv_j (\a) \epc \qd
     \ph (\la_j) = \tilde \ph (\la_j; 0) \epp
\end{equation}
Since only the residua at the simple poles at $\x_j^2$ enter the
formula for the correlation functions, there is a certain degree
of arbitrariness in the definition of the operator $\hv (\z, \a)$.

Equations (\ref{om1resform}) and (\ref{om2resform}) can be used
in order to calculate short-distance correlators explicitly on a
computer. Examples will be shown in section \ref{sec:examples}
below. The problem clearly divides into two separate tasks. The
first task is the calculation of $\Om_{i,j}^\pm$ and $\hv_j$.
For this purpose one has to generate the operators $\kv (\z, \a)$
and $\hv (\z, \a)$ for a finite number of lattice sites and then
calculate their residua. The second task is the actual calculation
of the functions $\om (\m_1, \m_2)$, $\om' (\m_1, \m_2)$ and
$\ph (\m)$. They will be described in the following section.

Still, the knowledge of the operators $\Om_1$ and $\Om_2$ in the
form (\ref{om1resform}) and (\ref{om2resform}) would not be
very useful, if these operators would not be nilpotent, which causes
the exponential series in (\ref{Omega1and2}) to terminate after
finitely many terms. As was first shown in \cite{BJMST04b} we have
\begin{subequations}
\begin{align}
      & \Om_{i,j}^+ \Om_{k,l}^+ = \Om_{i,j}^+ \Om_{k,l}^- =
         \Om_{i,j}^- \Om_{k,l}^- = 0 \epc
         \qd \text{if $\{i, j\} \cap \{k, l\} \ne \emptyset$} \epc \\[1ex]
      & [\Om_{i,j}^+, \Om_{k,l}^+] = [\Om_{i,j}^+, \Om_{k,l}^-] =
        [\Om_{i,j}^-, \Om_{k,l}^-] = 0 \epc
         \qd \text{if $\{i, j\} \cap \{k, l\} = \emptyset$}
\end{align}
\end{subequations}
which is a consequence of the fermionic nature of the operators
$\bv (\z, \a)$ and $\cv (\z, \a)$. It follows for chain segments
of length $n$ that
\begin{equation}
      \Om_1^{\left[\frac{n}{2}\right] + 1} = 0 \epc
\end{equation}
where $[x]$ stands for the integer part of $x$. Again for
$\Om_2$ our knowledge is so far restricted to what be
learned from explicit examples. We calculated $\Om^+_{i,j}$,
$\Om^-_{i,j}$ and $\hv_j$ explicitly in the inhomogeneous case for
$n = 1, 2, 3, 4$. The reader might wonder why we do not
present the matrices here. The reason is that they are very
big. Already for $n = 4$ they are so big that they can not be
handled any longer by an all-purpose computer algebra program
like Mathematica. We used the program FORM \cite{Vermaseren00}
instead which turned out to be efficient. We found that for $n$
up to 4 the $\hv_j$ mutually anticommute
\begin{equation} \label{hanti}
     \hv_j \hv_k + \hv_k \hv_j = 0 
\end{equation}
and that they commute with the $\Om_{i,j}^\pm$,
\begin{equation} \label{homcom}
     [\Om_{i,j}^\pm, \hv_k] = 0 \epp
\end{equation}
We also have the stronger property
\begin{equation} \label{hannihilom}
     \Om_{i,j}^\pm \hv_k = \hv_k \Om_{i,j}^\pm = 0 \qd
        \text{if $k \in \{i, j\}$.}
\end{equation}

We conjecture that (\ref{hanti})-(\ref{hannihilom}) are true in
general. This would mean, in particular, that
\begin{equation}
     [\Om_1, \Om_2] = 0
\end{equation}
and that $\Om_2$ is nilpotent,
\begin{equation}
     \Om_2^2 = 0 \epp
\end{equation}
As a further consequence the exponential form (\ref{Omega1and2})
would factorize,
\begin{equation}
     \re^{\Om_1 + \Om_2} =
        \re^{\Om_1} \re^{\Om_2} = \re^{\Om_1}(1 + \Om_2) =
	(1 + \Om_2) \re^{\Om_1} \epp
\end{equation}
We verified this explicitly for $n = 1, 2, 3, 4$.

\section{The physical part of the construction} \label{sec:physpart}
As we have seen above we need the functions $\om (\m_1, \m_2)$,
$\om' (\m_1, \m_2)$ and $\ph (\m)$ for the description of the
physical correlation functions in the limit $\alpha \rightarrow 0$.
A relatively simple description of these functions in terms of
an auxiliary function $\fa$ and a generalized magnetization density
$G$ was given in \cite{BGKS07}. We call it the $\fa$-formulation.
The $\fa$-formulation is useful for deriving multiple integral
formulae \cite{GKS04a} and for studying the high-temperature expansion
of the free energy and the correlation functions \cite{TsSh05}. For
the purpose of the actual numerical calculation of thermodynamic
properties \cite{Kluemper92,Kluemper93} or short-distance correlators
from the multiple integral \cite{BoGo05} one has to resort to a
different formulation we refer to as the $\fb \fbq$-formulation.
It needs pairs of functions, but the defining integral equations
in the massless regime $|\D| < 1$ involve only the real axis as
integration contour and become trivial in the zero temperature limit.
It turns out that the $\fb \fbq$-formulation is numerically extremely
stable (see below).

Switching from the $\fa$- to the $\fb \fbq$-formulation is a rather
standard procedure which basically requires the application of the
Fourier transformation (see e.g.\ \cite{BoGo05}). For this reason we
present only the result. We restrict ourselves to the massless
regime $|\D| < 1$ and set $\g = - \i \h \in {\mathbb R}$. Let us
define a pair of auxiliary functions as the solution of the non-linear
integral equations
\begin{subequations}
\label{nlies}
\begin{align}
     \ln \fb (x) & = - \frac{\p h}{2 (\p - \g) T}
                     - \frac{2 \p J \sin (\g)}{T \g \ch (\p x/\g)}
                     + \int_{- \infty}^\infty \frac{\rd y}{2 \p}
                       F(x - y) \ln (1 + \fb (y)) \notag \\
                     & \mspace{215.mu}
                     - \int_{- \infty}^\infty \frac{\rd y}{2 \p}
                       F(x - y + \h^-) \ln (1 + \fbq (y)) \epc \\[2ex]
     \ln \fbq (x) & = \frac{\p h}{2 (\p - \g) T}
                      - \frac{2 \p J \sin (\g)}{T \g \ch (\p x/\g)}
                     + \int_{- \infty}^\infty \frac{\rd y}{2 \p}
                       F(x - y) \ln (1 + \fbq (y)) \notag \\
                     & \mspace{215.mu}
                     - \int_{- \infty}^\infty \frac{\rd y}{2 \p}
                       F(x - y - \h^-) \ln (1 + \fb (y))
\end{align}
\end{subequations}
with kernel
\begin{equation}
     F(x) = \int_{- \infty}^\infty \rd k
            \frac{\sh \bigl( (\frac{\p}{2} - \g)k \bigr) \re^{\i kx}}
                 {2 \sh \bigl( (\p - \g)\frac{k}{2} \bigr)
                  \ch \bigl( \frac{\g k}{2} \bigr)} \epp
\end{equation}
Note that the physical parameters temperature $T$, magnetic field $h$,
and coupling $J$ enter only through the driving terms of equations
(\ref{nlies}) into our formulae. In this formulation (\ref{nlies}) is
valid for $0 < \g \le \p/2$ which means $0 \le \D < 1$. In order to
access negative $\D$ we have to reverse the sign of $J$ and conjugate
the local operators whose correlation functions we are interested in
by $\s^z$ on every second site of the spin chain (see \cite{YaYa66a}).

Except for the auxiliary functions $\fb$ and $\fbq$ we need two more pairs
of functions $g_\m^{(\pm)}$ and ${g'}_\m^{(\pm)}$ in order to define
$\om$ and $\om'$. Both pairs satisfy linear integral equations
involving $\fb$ and $\fbq$,
\begin{subequations}
\label{glies}
\begin{align}
     g_\m^{(+)} (x) & = \tst{\frac{\i \p}{\g}}
        \sech \Bigl( \tst{\frac{\p (x - \m)}{\g}} \Bigr)
        \notag \\ & \mspace{35.mu}
        + \int_{- \infty}^\infty \frac{\rd y}{2 \p}
          \frac{F(x - y)}{1 + \fb^{-1} (y)} g_\m^{(+)} (y)
        - \int_{- \infty}^\infty \frac{\rd y}{2 \p}
          \frac{F(x - y + \h^-)}{1 + \fbq^{-1} (y)}
          g_\m^{(-)} (y) \epc \displaybreak[0] \\[2ex]
     g_\m^{(-)} (x) & = \tst{\frac{\i \p}{\g}}
        \sech \Bigl( \tst{\frac{\p (x - \m)}{\g}} \Bigr)
        \notag \\ & \mspace{35.mu}
        + \int_{- \infty}^\infty \frac{\rd y}{2 \p}
          \frac{F(x - y)}{1 + \fbq^{-1} (y)}
          g_\m^{(-)} (y)
        - \int_{- \infty}^\infty \frac{\rd y}{2 \p}
          \frac{F(x - y - \h^-)}{1 + \fb^{-1} (y)} g_\m^{(+)} (y)
\end{align}
\end{subequations}
and
\begin{subequations}
\label{gplies}
\begin{align}
     {g'}_\m^{(+)} (x) & =
         \Bigl( \tst{\frac{\i \p}{\g} (x - \m) - \frac{\p}{2}} \Bigr)
         \sech \Bigl( \tst{\frac{\p (x - \m)}{\g}} \Bigr)
         \notag \\ & \mspace{20.mu}
         + \g \int_{- \infty}^\infty \frac{\rd y}{2 \p}
           \frac{D(x - y)}{1 + \fb^{-1} (y)} g_\m^{(+)} (y)
         - \g \int_{- \infty}^\infty \frac{\rd y}{2 \p}
           \frac{D(x - y + \h^-)}{1 + \fbq^{-1} (y)} g_\m^{(-)} (y)
         \notag \\ & \mspace{20.mu}
         + \int_{- \infty}^\infty \frac{\rd y}{2 \p}
           \frac{F(x - y)}{1 + \fb^{-1} (y)} {g'}_\m^{(+)} (y)
         - \int_{- \infty}^\infty \frac{\rd y}{2 \p}
           \frac{F(x - y + \h^-)}{1 + \fbq^{-1} (y)} {g'}_\m^{(-)} (y)
           \epc \displaybreak[0] \\[2ex]
     {g'}_\m^{(-)} (x) & =
         \Bigl( \tst{\frac{\i \p}{\g} (x - \m) + \frac{\p}{2}} \Bigr)
         \sech \Bigl( \tst{\frac{\p (x - \m)}{\g}} \Bigr)
         \notag \\ & \mspace{20.mu}
         + \g \int_{- \infty}^\infty \frac{\rd y}{2 \p}
           \frac{D(x - y)}{1 + \fbq^{-1} (y)} g_\m^{(-)} (y)
         - \g \int_{- \infty}^\infty \frac{\rd y}{2 \p}
           \frac{D(x - y - \h^-)}{1 + \fb^{-1} (y)} g_\m^{(+)} (y)
         \notag \\ & \mspace{20.mu}
         + \int_{- \infty}^\infty \frac{\rd y}{2 \p}
           \frac{F(x - y)}{1 + \fbq^{-1} (y)} {g'}_\m^{(-)} (y)
         - \int_{- \infty}^\infty \frac{\rd y}{2 \p}
           \frac{F(x - y - \h^-)}{1 + \fb^{-1} (y)} {g'}_\m^{(+)} (y)
           \epc
\end{align}
\end{subequations}
where
\begin{equation}
     D(x) = \int_{- \infty}^\infty \rd k
            \frac{\sin(kx) \sh \bigl( \frac{\p k}{2} \bigr)
                  \ch \bigl( (\frac{\p}{2} - \g)k \bigr)}
                 {4 \sh^2 \bigl( (\p - \g)\frac{k}{2} \bigr)
                  \ch^2 \bigl( \frac{\g k}{2} \bigr)} \epp
\end{equation}

The functions $\om (\m_1, \m_2)$, $\om' (\m_1, \m_2)$ and $\ph (\m)$
that determine the explicit form of the inhomogeneous correlation
functions of the XXZ chain can be written as integrals involving
$\fb$, $\fbq$, $g_\m^{(\pm)}$ and ${g'}_\m^{(\pm)}$. The function
\begin{equation}
     \ph(\m) = \int_{- \infty}^\infty \frac{\rd x}{2 (\p - \g) \i}
               \biggl[ \frac{g_\m^{(+)} (x)}{1 + \fb^{-1} (x)}
                     - \frac{g_\m^{(-)} (x)}{1 + \fbq^{-1} (x)} \biggl]
               \epc
\end{equation}
also determines the magnetization
\begin{equation}
     m(T,h) = - \tst{\2} \ph (0)
\end{equation}
which is the only independent one-point function of the XXZ chain.
The function
\begin{multline}
     \om(\m_1, \m_2) =
        - \tst{\2} K(\m_1 - \m_2) - \int_{- \infty}^\infty \rd k 
          \frac{\sh \bigl( (\p - \g)\frac{k}{2} \bigr)
                \cos(k(\m_1 - \m_2))}
               {\i \sh \bigl( \frac{\p k}{2} \bigr)
                   \ch \bigl( \frac{\g k}{2} \bigr)} \\[1ex]
             - \int_{- \infty}^\infty \frac{\rd x}{\g}
               \sech \Bigl( \tst{\frac{\p (x - \m_2)}{\g}} \Bigr)
               \biggl[ \frac{g_{\m_1}^{(+)} (x)}{1 + \fb^{-1} (x)}
                     + \frac{g_{\m_1}^{(-)} (x)}{1 + \fbq^{-1} (x)}
               \biggl]
\end{multline}
with
\begin{equation}
     K(\m) = \cth(\m - \h) - \cth(\m + \h)
\end{equation}
also determines the internal energy
\begin{equation}
     e(T,h) = J \< \s_{j-1}^x \s_j^x
              + \s_{j-1}^y \s_j^y + \D (\s_{j-1}^z \s_j^z - 1) \>_{T, h}
            = J \bigl( \sh(\h) \om(0,0) - \D \bigr)
\end{equation}
of the XXZ chain. The function $\om' (\m_1, \m_2)$ is defined as
\begin{multline}
     \om'(\m_1, \m_2) =
        \tst{\frac{\h}{2}} K^{(+)} (\m_1 - \m_2)
        + \int_{- \infty}^\infty \rd k 
          \frac{\g \sin(k(\m_1 - \m_2))}
               {2 \i \tgh \bigl( \frac{\p k}{2} \bigr)
                   \ch^2 \bigl( \frac{\g k}{2} \bigr)} \\[1ex]
             - \int_{- \infty}^\infty \frac{\rd x}{\g}
               \sech \Bigl( \tst{\frac{\p (x - \m_2)}{\g}} \Bigr)
               \biggl[ \frac{f_{\m_1}^{(+)} (x)}{1 + \fb^{-1} (x)}
                     + \frac{f_{\m_1}^{(-)} (x)}{1 + \fbq^{-1} (x)}
               \biggl] \\[1ex]
             + \int_{- \infty}^\infty \frac{\rd x}{\g} (x - \m_2)
               \sech \Bigl( \tst{\frac{\p (x - \m_2)}{\g}} \Bigr)
               \biggl[ \frac{g_{\m_1}^{(+)} (x)}{1 + \fb^{-1} (x)}
                     + \frac{g_{\m_1}^{(-)} (x)}{1 + \fbq^{-1} (x)}
               \biggl] \epc
\end{multline}
where
\begin{equation}
     K^{(+)} (\m) = \cth(\m - \h) + \cth(\m + \h) \epc \qd
     f_\m^{(\pm)} (x) = {g'}_\m^{(\pm)} (x)
                        \mp \tst{\frac{\i \g}{2}} g_\m^{(\pm)} (x) \epp
\end{equation}

From the symmetry $\om (\m_1, \m_2, \a) = \om (\m_2, \m_1, -\a)$
(see \cite{BGKS07}) we conclude that
\begin{equation}
     \om (\m_1, \m_2) = \om (\m_2, \m_1) \epc \qd
     \om' (\m_1, \m_2) = - \om' (\m_2, \m_1) \epp
\end{equation}
This property was crucial for the derivation of (\ref{om1resform}).
It also implies that $\om' (0,0) = 0$. The first derivate of
this function with respect to the spectral parameter,
$\om'_x = \6_\m \om'(\m,0)|_{\m = 0}$,  appears in the homogeneous
density matrix $D_2 (T,h)$ and, hence, is related to the neighbour
correlators (see \cite{BGKS07}).

For the calculation of the physical correlation functions one
has to perform the homogeneous limit. This limit is rather singular,
because the coefficients coming from the algebraic part in general
have poles when two inhomogeneities coincide. These poles are canceled
by zeros stemming from certain symmetric combinations of the functions
$\om (\la_1, \la_2)$, $\om' (\la_1, \la_2)$ and $\ph (\la)$ in the
numerators. In order to perform the limit one has to apply l'H\^ospital's
rule, finally leading to polynomials in the functions $\om(0,0)$,
$\om'(0,0)$, $\ph(0)$ and derivatives of these functions with respect
to the inhomogeneity parameters evaluated at zero. We shall denote
derivatives with respect to the first argument by subscripts $x$
and derivatives with respect to the second argument by subscripts
$y$ and leave out zero arguments for simplicity. Then e.g.\ $\om'_{xyy} =
\6_x \6_y^2 \om' (x,y)|_{x,y = 0}$ etc.

For the examples in the next section the non-linear integral
equations for $\fb$ and $\fbq$ as well as their linear counterparts
for $g_\m^{(\pm)}$ and ${g'}_\m^{(\pm)}$ were solved iteratively in
Fourier space, utilizing frequently the fast Fourier transformation
algorithm. The derivatives of $g_\m^{(\pm)}$ and ${g'}_\m^{(\pm)}$
with respect to $\m$, needed in the computation of the respective
derivatives of $\om$, $\om'$ and $\ph$, satisfy linear integral
equations as well, which were obtained as derivatives of the
equations for $g_\m^{(\pm)}$ and ${g'}_\m^{(\pm)}$. Taking into
account derivatives is particularly simple in Fourier space.

\section{Examples of short-distance correlators} \label{sec:examples}
Here we mostly concentrate on the longitudinal and transversal
two-point functions $\<\s_1^z \s_n^z\>_{T,h}$, $\<\s_1^x \s_n^x\>_{T, h}$
for $n = 2, 3, 4$. The algebraic part for $n = 2, 3$ was already
calculated in the more general inhomogeneous case in our previous
paper \cite{BGKS07}.

In the inhomogeneous case for $n = 2$ we obtained
\begin{multline}
     \4 \tr_{12} \bigl( \re^{\Om_1}(1 + \Om_2)\s_1^z \s_2^z \bigr)
        = \4 \tr_{12} \bigl( \Om_1 \s_1^z \s_2^z \bigr) \\
        = - \4 \tr_{12} \bigl( [\om (\la_1, \la_2) \Om_{1,2}^+
	       + \om' (\la_1, \la_2) \Om_{1,2}^-] \s_1^z \s_2^z \bigr) \\
        = \cth (\h) \om(\la_1, \la_2)
	  + \frac{\cth(\la_1 - \la_2)}{\h} \om'(\la_1, \la_2)
\end{multline}
and in the homogeneous limit $\la_j \rightarrow 0$,
\begin{equation}
     \<\s_1^z \s_2^z\>_{T, h} = \cth (\h) \om + \frac{\om'_x}{\h} \epp
\end{equation}
Note that the limit exists, because $\om'(\la_1, \la_2)$ is antisymmetric.
Similarly for the transversal correlation functions,
\begin{multline}
     \4 \tr_{12} \bigl( \re^{\Om_1}(1 + \Om_2)\s_1^x \s_2^x \bigr)
        = \4 \tr_{12} \bigl( \Om_1 \s_1^x \s_2^x \bigr) \\
        = - \4 \tr_{12} \bigl( [\om (\la_1, \la_2) \Om_{1,2}^+
	       + \om' (\la_1, \la_2) \Om_{1,2}^-] \s_1^x \s_2^x \bigr) \\
        = - \frac{\ch (\la_1 - \la_2)}{2 \sh(\h)} \om(\la_1, \la_2)
	  - \frac{\ch(\h)}{2 \h \sh(\la_1 - \la_2)} \om'(\la_1, \la_2)
\end{multline}
which in the homogenous limit becomes
\begin{equation}
     \<\s_1^x \s_2^x\>_{T, h} =
        - \frac{\om}{2 \sh(\h)} - \frac{\ch(\h) \om'_x}{2 \h} \epp
\end{equation}

The two-point functions $\<\s_1^z \s_n^z\>_{T,h}$,
$\<\s_1^x \s_n^x\>_{T, h}$ are even under spin reversal. Therefore
we do not expect any non-vanishing contribution from $\Om_2$,
which would come with functions $\ph$, i.e.\ odd functions of
the magnetic field $h$. This is indeed what we observe from our
calculation for $n = 1, 2, 3, 4$. The above two-point correlation
functions depend only on $\om$, $\om'$ and their derivatives.
As an example of a correlation function which is of no definite
symmetry under spin reversal and which does depend on $\ph$ and
its derivatives we shall discuss the emptiness formation probability
below.

We obtained the formula for $n = 3, 4$ in the same way as for $n = 2$
by first calculating the inhomogeneous generalization of the
correlation function and performing the homogeneous limit at
the end. The formula for $\<\s_1^z \s_3^z\>_{T,h}$ in the inhomogeneous
case was presented in \cite{BGKS07}. For space limitations we do
not repeat it here and also do not show the inhomogeneous formulae
in the remaining cases. Instead we merely list the results
for the physical correlation functions. For $n = 3$ we already
obtained in \cite{BGKS07}
\begin{subequations}
\begin{align}
     & \<\s^z_1 \s^z_3\>_{T, h} =
        2 \cth(2\h) \omega
        + \frac{\omega^{\prime}_x}{\eta}
	+ \frac{\tgh(\h) (\omega_{xx} - 2 \omega_{xy})}{4}
        - \frac{\sh^2(\h) \omega^{\prime}_{xxy}}{4 \eta} \epc \\[1ex]
     & \<\s^x_1 \s^x_3 \>_{T,h} =
        - \frac{1}{\sh(2 \h)} \omega
        - \frac{\ch(2\h)}{2\h} \omega^{\prime}_x
	  \notag \\ & \mspace{180.mu}
	- \frac{\ch(2\h) \tgh(\h) (\omega_{xx} - 2 \omega_{xy})}{8}
	+ \frac{\sh^2(\h) \omega^{\prime}_{xxy}}{8\h} \epp
\end{align}
\end{subequations}
The length of the formulae grows rapidly with the number of lattice
sites. For $n = 4$ we found
\begin{multline} \label{szsz4}
     \<\s_1^z \s_4^z\>_{T, h} =
     \frac{1}{768 q^4 \left(- 1 + q^6 \right) \left(1+q^2\right)
              \eta ^2} \\[1ex]
     \biggl\{ 384 q^4 \left(1+q^2\right)^2
                    \left(5-4 q^2+5 q^4\right) \eta ^2 \om
		    \mspace{210.mu}\\
     -8 \left(1+q^4 \left(52+64 q^2-234 q^4+64 q^6+52 q^8
               +q^{12}\right)\right) \eta ^2 \om_{xy} \\
     + 192 q^4 \left(-1+q^2\right)^2
        \left(1+4 q^2+q^4\right) \eta ^2 \om_{yy} \\
     + \left(-1+q^2\right)^4 \left(1+q^4\right) \left(1+4 q^2+q^4\right)
       \eta^2 \Bigl[ -4 \om_{xyyy} + 6 \om_{xxyy} \Bigr] \\[1ex]
     - 768 q^4 \left(-1-q^2+q^6+q^8\right) \eta {\om'}_y \\
     +16 \left(-1+q^2\right)^3
         \left(1+6 q^2+11 q^4+11 q^6+6 q^8+q^{10}\right)
         \eta  {\om'}_{xyy}\\
     -2 \left(-1+q^2\right)^5 \left(1+2 q^2+2 q^4+q^6\right)
        \eta  {\om'}_{xxyyy} \displaybreak[0] \\
     + 8 \left(-1+q^2\right)^3 \left(1+q^2\right)
         \left(1+6 q^2+34 q^4+6 q^6+q^8\right) \eta ^2
	 \Bigl[\om_y^2 - \om \om_{xy} \Bigr] \\
     + \left(-1-4 q^2-22 q^4-12 q^6+12 q^{10}+22 q^{12}
             +4 q^{14}+q^{16}\right) \eta ^2
     \Bigl[ - 6 \om_{yy}^2 \\
     +12 \om_{yy} \om_{xy} + 4 \om_y \om_{yyy}
     -12 \om_y \om_{xyy} -4  \om \om_{xyyy}
     +6  \om \om_{xxyy} \Bigr] \displaybreak[0] \\
     + 16 \left(-1+q^2\right)^4 \left(1+q^2\right)^2
          \left(1+q^2+q^4\right) \eta
     \Bigl[ \om_{yy} {\om'}_y - \om_y {\om'}_{yy} + \om {\om'}_{xyy}
     \Bigr] \\
     + \left(-1+q^4\right)^2 \left(1+5 q^2+6 q^4+5 q^6+q^8\right) \eta
       \Bigl[
     4 \om_{xyyy} {\om'}_y -6 \om_{xxyy} {\om'}_y
     -2 \om_{yyy} {\om'}_{yy} \\
     +6 \om_{xyy} {\om'}_{yy} +2 \om_{yy} {\om'}_{yyy}
     -4 \om_{xy} {\om'}_{yyy} -6 \om_{yy} {\om'}_{xyy}
     +4 \om_y {\om'}_{xyyy} -2 \om {\om'}_{xxyyy}\Bigr]\\
     + 3 \left(-1+q^4\right)^3 \left(1+q^2+q^4\right)
     \Bigl[ {\om'}_{yyy} {\om'}_{xyy} - {\om'}_{yy} {\om'}_{xyyy}
     + {\om'}_y {\om'}_{xxyyy} \Bigr] \biggr\}
\end{multline}
and in the transversal case,
\begin{multline} \label{sxsx4}
     \<\s_1^x \s_4^x\>_{T, h} =
     \frac{1}{3072 q^5
     \left(-1+q^6\right) \eta ^2} \\
     \biggr\{
     -768 q^6 \left(1+10 q^2+q^4\right) \eta^2 \om \mspace{270.mu} \\
     +16 q^2 \left(-1+q^2\right)^2
        \left(31+56 q^2-30 q^4+56 q^6+31 q^8\right) \eta ^2 \om_{xy}\\
     -96 q^2 \left(-1+q^2\right)^2
      \left(3+5 q^2-4 q^4+5 q^6+3 q^8\right) \eta ^2 \om_{yy} \\
     + q^2 \left(-1+q^2\right)^4 \left(1+4 q^2+q^4\right)
     \eta^2 \Bigl[ 8 \om_{xyyy} -12 \om_{xxyy} \Bigr]\\
     + 192 q^2 \left(-3-q^2-q^4+q^8+q^{10}+3 q^{12}\right) \eta {\om'}_y \\
     +8 \left(-1+q^2\right)^3
        \left(1-12 q^2-25 q^4-25 q^6-12 q^8+q^{10}\right)
	\eta  {\om'}_{xyy}\\
     +2 \left(-1+q^2\right)^5 \left(1+2 q^2+2 q^4+q^6\right)
        \eta {\om'}_{xxyyy} \displaybreak[0] \\
     +16 q^2 \left(-1+q^2\right)^3 \left(17+7 q^2+7 q^4+17 q^6\right)
        \eta^2 \Bigl[ \om \om_{xy} - \om_y^2 \Bigr] \\
     + q^2 \left(-5-4 q^2-13 q^4+13 q^8+4 q^{10}+5 q^{12}\right) \eta ^2
     \Bigl[ 12 \om_{yy}^2 -24 \om_{yy} \om_{xy} \\
     -8 \om_y \om_{yyy} +24 \om_y \om_{xyy} +8 \om \om_{xyyy}
     -12 \om \om_{xxyy} \Bigr]\\
     + 8 \left(-1+q^2\right)^4 \left(1-9 q^2-8 q^4-9 q^6+q^8\right) \eta
     \Bigr[ \om_{yy} {\om'}_y - \om_y {\om'}_{yy}
           + \om {\om'}_{xyy} \Bigr] \displaybreak[0] \\
     + \left(-1+q^4\right)^2 \left(1+5 q^2+6 q^4+5 q^6+q^8\right) \eta
     \Bigl[ -4 \om_{xyyy} {\om'}_y +6 \om_{xxyy} {\om'}_y\\
     +2 \om_{yyy} {\om'}_{yy} -6 \om_{xyy} {\om'}_{yy}
     -2 \om_{yy} {\om'}_{yyy} +4 \om_{xy} {\om'}_{yyy}
     +6 \om_{yy} {\om'}_{xyy} -4 \om_y {\om'}_{xyyy}
     +2 \om {\om'}_{xxyyy} \Bigr] \displaybreak[0] \\
     +3 \left(-1+q^4\right)^3 \left(1+q^2+q^4\right)
     \Bigl[- {\om'}_{yyy} {\om'}_{xyy} + {\om'}_{yy} {\om'}_{xyyy}
     - {\om'}_y {\om'}_{xxyyy} \Bigr] \biggl\} \epp
\end{multline}

\begin{figure}
    \centering
    \includegraphics{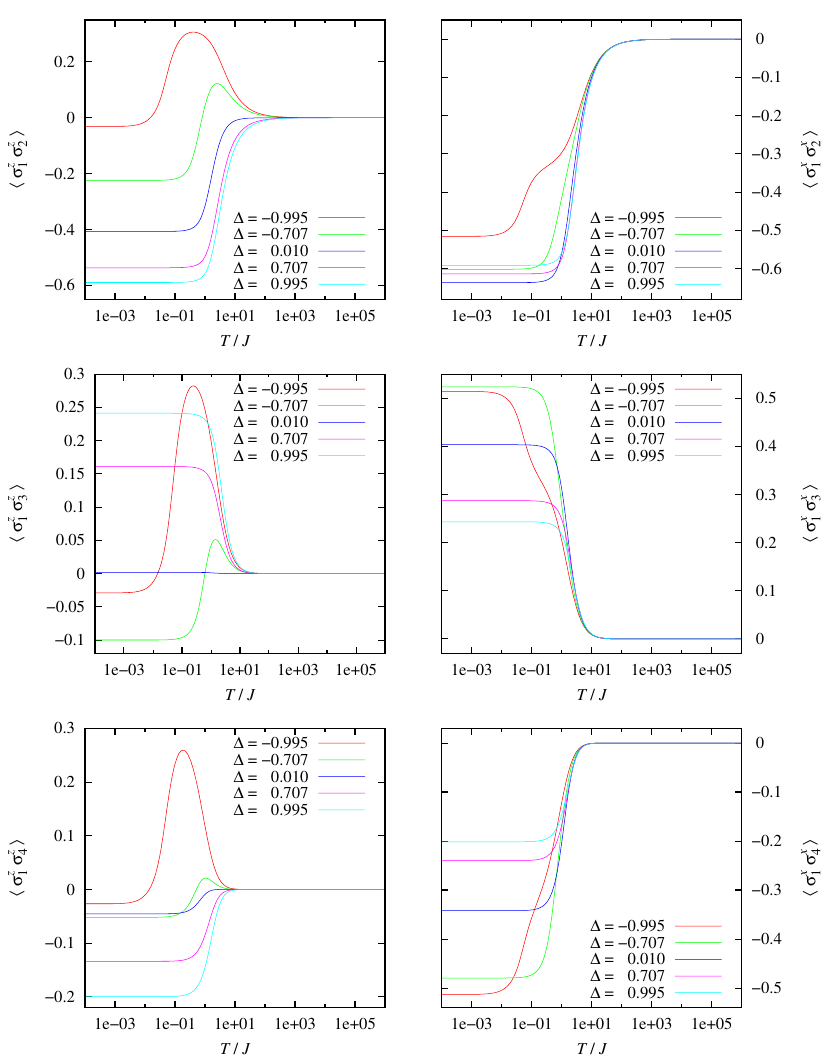}
    \caption{\label{fig:tvardelta} Two-point correlators for $n= 2, 3, 4$
             as a function of temperature for various values of $\D$ in
	     the massless regime at zero magnetic field.} 
\end{figure}  

\begin{figure}
    \centering
    \includegraphics{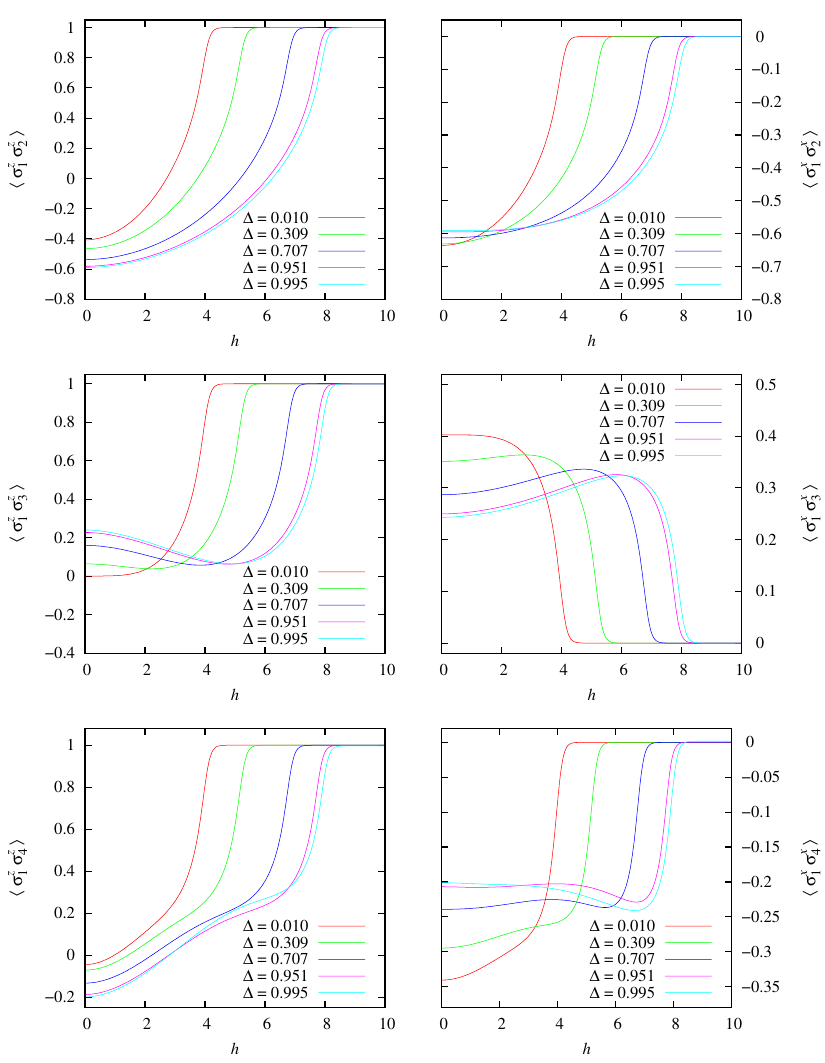}
    \caption{\label{fig:hvardelta} Two-point correlators for $n= 2, 3, 4$
             and $T/J = 0.1$ as a function of magnetic field for
	     various values of $\D$.} 
\end{figure}

Using the representations of the previous section for the functions
$\om$ and $\om'$ we can determine high-precision numerical values for
the various two-point correlators. Figures \ref{fig:tvardelta}-%
\ref{fig:hvardelta} show selected examples. For $n = 2, 3, 4$ the
longitudinal correlation functions $\<\s_1^z \s_n^z\>_{T, h}$ are
exhibited in the left panels, while the right panels show the transversal
correlation functions $\<\s_1^x \s_n^x\>_{T, h}$. We observe a rich
scenario for their  temperature and field dependence, in particular
non-monotonous behaviour at intermediate temperatures. The numerical
precision in all figures is between 3 and 4 significant digits.

In figure \ref{fig:tvardelta} we display our data for vanishing magnetic
field and various values of $\D \in (-1,1)$. The transversal correlations
are alternating in sign with $n$ for all values of $\D$. But for the
longitudinal correlations their qualitative behaviour depends on the
sign of $\D$. They alternate for positive $\D$. For negative $\D$ they
are always negative at sufficiently low temperatures, change their sign at
a `crossover temperature' $T_0 (n;\D)$, reach a positive maximum and
tend to zero from above in the high temperature limit. This phenomenon
was studied numerically on finite chains of 16 and 18 sites in
\cite{FaMc99} and was interpreted there as a `quantum-to-classical
crossover'. In \cite{FaMc99} it was estimated that the crossover
temperature $T_0 (n;\D)$ remains finite in the thermodynamic limit.
From our formulae we can calculate it with high precision (see table
\ref{tab:famc}).

\begin{table}[t]
\begin{minipage}{\linewidth}
\renewcommand{\thefootnote}{\thempfootnote}
    \centering
    \begin{tabular}{>{$}c<{$}*{3}{>{$}c<{$}>{($}c<{$)}}}
      \toprule
        \D & n = 2 & L = 18 & n = 3 & L = 18 & n = 4 & L = 18 \\
      \midrule
	-0.1 & 4.96645  & 4.966 & 3.32288  & 3.323 & 2.56077  & 2.561 \\
	-0.2 & 2.43157  & 2.432 & 1.64332  & 1.643 & 1.27520  & 1.275 \\
	-0.3 & 1.56079  & 1.561 & 1.07081  & 1.071 & 0.839169 & 0.839 \\
	-0.4 & 1.10294  & 1.103 & 0.771287 & 0.771 & 0.611558 & 0.612 \\
	-0.5 & 0.806967 & 0.807 & 0.577718 & 0.578 & 0.4641%
	 \footnote{The prefactor on the rhs of equation (\ref{szsz4})
	           has a pole as a function of $q$ when $q$ is a third
		   root of unity corresponding to $\D = - 0.5$. We
		   estimated the value of $T_0 (n;\D)$ for $\D = - 0.5$
		   by interpolating between two values closely above
		   and below $-0.5$.} & 0.464 \\
	-0.6 & 0.588818 & 0.589 & 0.434179 & 0.434 & 0.354030 & 0.355 \\
	-0.7 & 0.412795 & 0.413 & 0.316321 & 0.318 & 0.262606 & 0.264 \\
	-0.8 & 0.262355 & 0.265 & 0.211402 & 0.215 & 0.179803 & 0.184 \\
	-0.9 & 0.129195 & 0.137 & 0.111127 & 0.118 & 0.098055 & 0.104 \\
      \bottomrule
    \end{tabular}
    \caption{\label{tab:famc} Numerical values of the cross-over
             temperature $T_0 (n;\D)$ on the infinite chain and
	     for a chain of 18 sites in units of $J/2$ (values in
	     brackets, taken from \cite{FaMc99}).}
\end{minipage}
\end{table}

Note the peculiar low temperature behaviour close to $\D = - 1$,
where one would expect simultaneously isotropy ($\<\s^z_1 \s^z_n\> = 
(-1)^{n+1} \<\s^x_1 \s^x_n\>$) as well as full polarization
($\<\s^z_1 \s^z_n\> + 2 (-1)^{n+1} \<\s^x_1 \s^x_n\> = 1$). However,
this is not seen, not even for $\D$ as close to the isotropic point
\begin{figure}
    \centering
    \includegraphics{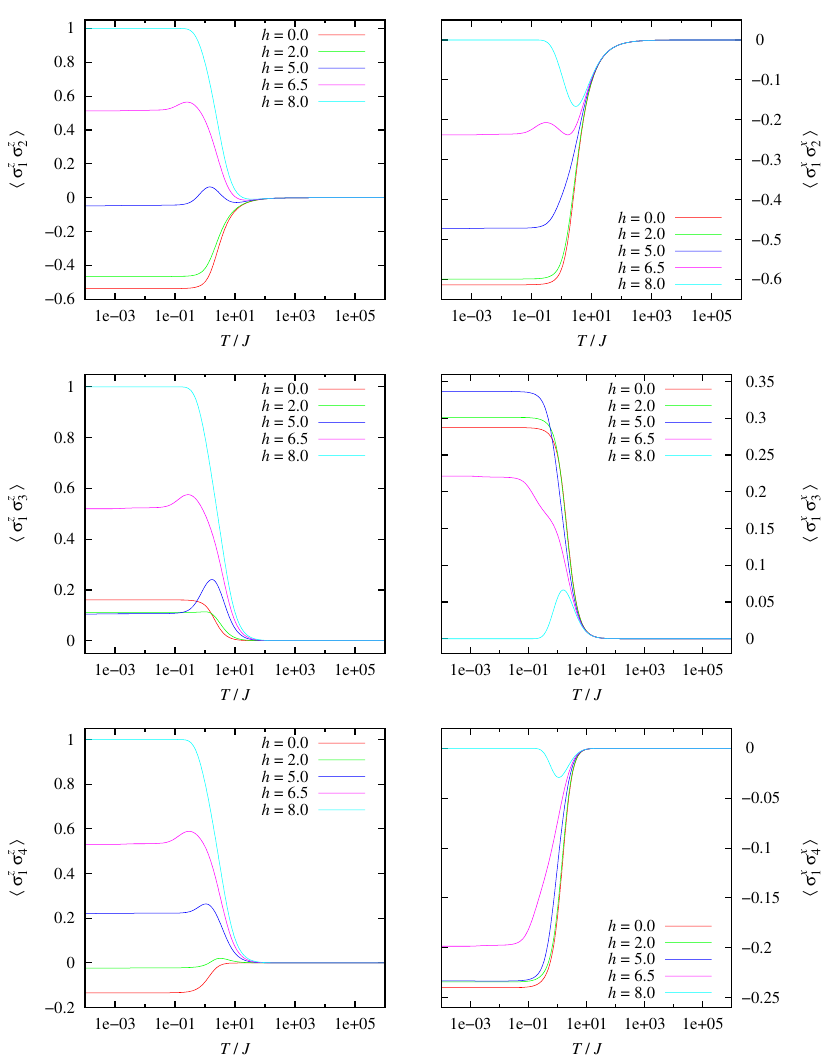}
    \caption{\label{fig:tvarh} Two-point correlators for $n= 2, 3, 4$
             and $\D = 1/\sqrt{2}$ as a function of temperature for
	     various magnetic fields.} 
\end{figure}  
as $- 0.995$, since the full polarization is still forced into the $xy$-%
plane by the slightly dominant transversal exchange. At elevated but
not too high temperature the system is still close to full polarization,
but the small anisotropy is irrelevant and longitudinal as well as
transversal correlation attain values close to $\pm 1/3$. At much higher
temperatures all correlations tend to zero. As was explained in
\cite{FaMc99} the negative sign of the longitudinal correlations
at low temperatures is a pure quantum effect.

In figure \ref{fig:tvarh} we plotted the correlators for a fixed
value of $\D = 1/\sqrt{2}$ for various magnetic fields. The curves
show non-monotonous behaviour for intermediate fields. In particular,
a sign change is possible for positive $\D$ if the magnetic field is
switched on, as can be seen in the upper and lower left panels.
In general, we observe a competition between the effect of the
magnetic field, which tends to align the spins in $z$-direction,
the longitudinal exchange interaction in the Hamiltonian, which
enforces antiparallel alignment of neighbouring spins, and the transversal
exchange interaction, which can be interpreted in terms of quantum
fluctuations. The temperature sets the scale for the relative and
absolute relevance of these terms. At low temperatures the quantum
fluctuations dominate, decreasing e.g.\ the value of $\<\s_1^z \s_3^z\>$
from 1 to a smaller positive value. When increasing the temperature, at
intermediate field the quantum fluctuations are suppressed and the
field is less hindered to align the spins, before at very high temperature
the thermal fluctuations destroy all order.

For large enough fields (light blue curves) the correlation functions
reach saturation. This can be also observed in figure~\ref{fig:hvardelta},
where the correlators are shown as functions of the magnetic field
for relatively low temperature, $T/J = 0.1$, and where the saturation
roughly sets in at its zero temperature value $h = (1 + \D) 4 J$.
An interesting feature is the decrease of $\<\s_1^z \s_3^z\>$ for small
fields which we interpret as due to a weakening of the antiferromagnetic
neighbour correlations.

As an independent test of our results, we numerically calculated the
correlation functions for finite systems. Since the Hilbert space
dimension grows exponentially with the chain length, this required us to
characterise the spectrum and eigenvectors of high-dimensional
sparse matrices. The complete diagonalisation of such matrices is
too time consuming or simply impossible due to memory limitations.
We therefore used Krylov space methods, in particular, Chebyshev
expansion~\cite{WWAF06}.
\begin{figure}
    \centering
    \includegraphics{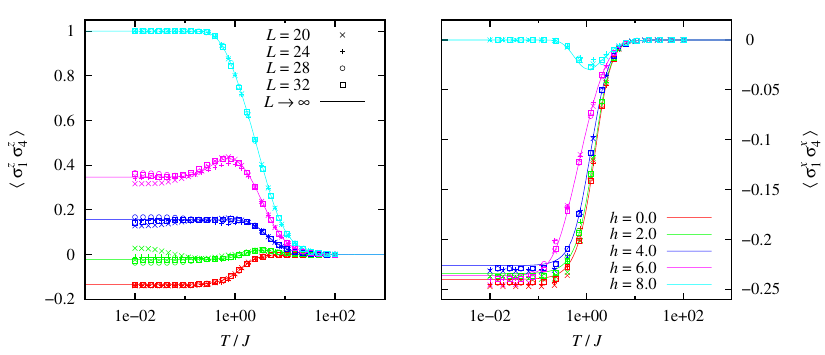}
    \caption{\label{fig:copare_ed} Two-point correlators for $n= 4$
             and $\D = 1/\sqrt{2}$ as a function of temperature,
	     comparison with data from direct diagonalization of finite
	     chains.} 
\end{figure}  

Given the $D$ eigenstates $|k\rangle$ of the system, we split the
expectation value of an operator $A$ into a low-energy and a
high-energy part,
\begin{align}
  Z & = \trace(\E^{-H/T}) 
      = \sum_{k=0}^{C-1} \E^{-E_k/T}
        + \sum_{k=C}^{D-1} \E^{-E_k/T}\,,\\
          \langle A\rangle & = \frac{1}{Z}\trace(A \E^{-H/T})
      = \frac{1}{Z}\left[\sum_{k=0}^{C-1} \langle k|A|k\rangle\,
        \E^{-E_k/T} +\sum_{k=C}^{D-1} \langle k|A|k\rangle\,
	\E^{-E_k/T} \right]\,,
\end{align}
where $H$ is the Hamiltonian of the system, $Z$ the partition
function, and $E_k$ the energy of the eigenstate $|k\rangle$. The $C$
low-energy states and the corresponding matrix elements can be
calculated exactly with the Lanczos algorithm~\cite{La50}. The
contribution of the remaining states is estimated via Chebyshev
expansions of the density of states,
\begin{align} \displaybreak[0]
  \sum_{k=C}^{D-1} &\E^{-E_k/T}  
    = \int\limits_{-\infty}^{\infty} \rho(E)\, \E^{-E/T}\,dE\,,\\
  \rho(E)
  & = \sum_{k=C}^{D-1} \delta(E-E_k)
    = \frac{1}{\pi\sqrt{s^2 - (E-\bar E)^2}} 
      \Big[\mu_0^\rho + 2\sum_{n=1}^{N-1} \mu_n^\rho
           T_n(\tfrac{E-\bar E}{s})\Big]\,,\\
      \mu_n^\rho & = g_n^N\trace(P T_n[(H-\bar E)/s])
	      \quad\text{with}\quad
  P = 1 - \sum_{k=0}^{C-1} |k\rangle\langle k|\,,
\end{align}
and of the diagonal matrix elements of $A$,
\begin{align}
  \sum_{k=C}^{D-1} &\langle k|A|k\rangle \E^{-E_k/T} 
    = \int\limits_{-\infty}^{\infty} a(E)\, \E^{-E/T}\,dE\,,\\ a(E)
  & = \sum_{k=C}^{D-1} \langle k|A|k\rangle\delta(E-E_k)
    = \frac{1}{\pi\sqrt{s^2 - (E-\bar E)^2}} 
      \Big[\mu_0^a + 2\sum_{n=1}^{N-1} \mu_n^a
           T_n(\tfrac{E-\bar E}{s})\Big]\,,\\
  \mu_n^a
  & = g_n^N\trace(P A T_n[(H-\bar E)/s])\,.
\end{align}
Here $T_n$ are the Chebyshev polynomials of the first kind and $\bar
E$, $s$ denote scaling factors that shift the spectrum of $H$ into the
domain of the polynomials $[-1,1]$. The coefficients $g_n^N$ account
for a special kernel~\cite{WWAF06,Ja12} which improves the convergence
of the truncated Chebyshev series.  The traces defining the moments
$\mu_n^\rho$ and $\mu_n^a$ can be estimated stochastically,
\begin{equation}
     \trace(X) \approx \tfrac{D-C}{R}\sum_{r=0}^{R-1} \langle r|X|r\rangle
        \quad\text{with}\quad R\ll D \text{ random states } |r\rangle\,.
\end{equation}
Thus the most time consuming part of the moment evaluation is the
recursive calculation of $T_n[(H-\bar E)/s]|r\rangle$ for a few dozen
states $|r\rangle$. The whole procedure is feasible for Hilbert space
dimensions of $10^7$ to $10^8$. Employing $S^z$ and momentum
conservation, with moderate effort we can reach a chain length of
$L=32$ (about 430~CPU hours on a 2.66~GHz Xeon Cluster).

We also performed tests against other exact results. We compared
the high temperature expansions of equations (\ref{szsz4}),
(\ref{sxsx4}) with the high temperature expansions for the multiple
integrals from \cite{GKS05} up to third order in $1/T$. We compared
with the known ground state results of \cite{KSTS03,KSTS04}. We compared
the $zz$-correlation functions in the XX limit with the explicit
result (see e.g.\ \cite{GoSe05}), and we computed the correlation
functions for $\D$ very close to 1 and compared with the results
obtained for the isotropic limit and $n = 2, 3$ in \cite{BGKS06}.

\begin{figure}
    \centering
    \includegraphics{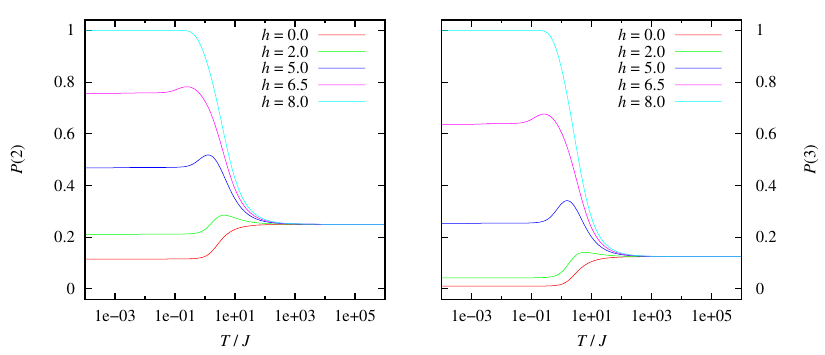}
    \caption{\label{fig:empt23} Emptiness formation probability
             for $n= 2, 3$ and $\D = 1/\sqrt{2}$ as a function of
	     temperature for various magnetic fields.} 
\end{figure}  

As an example of a correlation function which has no definite
symmetry with respect to the reversal of all spins and hence must
depend on $\ph$ and its derivatives we consider the emptiness
formation probability $P(n) = 2^{-n} \< \prod_{j=1}^n
(1 + \s_j^z)\>_{T, h}$. We know from \cite{BGKS07} that
\begin{equation}
     P(2) = \4 - \frac{\ph}{2} + \frac{\cth (\h) \om}{4}
            + \frac{\om_x'}{4 \h}
\end{equation}
and
\begin{multline}
     P(3) = \frac{1}{8}
	    - \frac{3 \ph}{8}
            + \frac{\left(1+q^2+q^4\right) \om}{2 \left(-1+q^4\right)}
	    + \frac{\left(-1+q^2\right) (\om_{yy} - 2 \om_{xy})}
	           {32 \left(1+q^2\right)}
	    - \frac{3 \om'_y}{8 \eta } \\
	    + \frac{\left(-1+q^2\right)^2 \om'_{xyy}}{128 q^2 \eta }
            + \frac{3 \left(1+q^2\right) (- \om \ph_{xx} + 2 \om_y \ph_x
	            + \om_{yy} \ph - 2 \om_{xy} \ph)}
	           {32 \left(-1+q^2\right)} \\
            + \frac{\left(1+10 q^2+q^4\right) (\om'_{xyy} \ph
                    - \om'_{yy} \ph_x + \om'_y \ph_{xx})}
	           {128 q^2 \eta }
	    \epp
\end{multline}
$P(2)$ and $P(3)$ are displayed in figure \ref{fig:empt23}. They show
the characteristic high temperature behaviour $P(n) \sim 2^{-n}$ and
saturate for small temperature and $h > (1 + \D) 4 J$. Like the
two-point functions they exhibit non-monotonous behaviour at
intermediate magnetic fields.

\section{Conclusion}
The aim of this work was to demonstrate that short-distance
correlation functions of the XXZ chain can be computed efficiently
from the exponential form of the density matrix proposed in
\cite{BGKS07}. We found that an accurate computation is possible
for arbitrary temperatures and magnetic fields in the massless
regime $- 1 < \D < 1$, even beyond the phase boundary to the
fully polarized state. This has to be contrasted with our experience
\cite{BoGo05} in computing the same correlation functions from the
multiple integral representations obtained in \cite{GKS04a,GKS05},
which appeared to be much harder. In fact, in \cite{BoGo05} we were
unable to proceed beyond $n = 3$. In the present approach the difficulty
for larger $n$ comes mainly from the algebraic part of the calculation.
We believe that ranges of the order of $n=10$ may be reached if
one works directly with the homogeneous transfer matrices (which would
mean to calculate higher order residua in (\ref{Omega1and2})).
\\[1ex]{\bf Acknowledgment.}
HB, JD and FG gratefully acknowledge financial support by the Volkswagen
Foundation and by the DFG through Graduiertenkolleg 1052. JS was
supported by a Grant-in-Aid for Scientific Research \#20540370 from the
Ministry of Education in Japan.


\begin{thebibliography}{10}

\bibitem{BGKS06}
H.~Boos, F.~G\"ohmann, A.~Kl\"umper, and J.~Suzuki, \emph{Factorization of
  multiple integrals representing the density matrix of a finite segment of the
  {H}eisenberg spin chain}, J. Stat. Mech. (2006), P04001.

\bibitem{BGKS07}
\bysame, \emph{Factorization of the finite temperature correlation functions of
  the {XXZ} chain in a magnetic field}, J. Phys. A \textbf{40} (2007), 10699.

\bibitem{BJMST04b}
H.~Boos, M.~Jimbo, T.~Miwa, F.~Smirnov, and Y.~Takeyama, \emph{Reduced $q${KZ}
  equation and correlation functions of the {XXZ} model}, Comm. Math. Phys.
  \textbf{261} (2006), 245.

\bibitem{BJMST07app}
\bysame, \emph{Fermionic basis for space of operators in the {XXZ} model},
  preprint, hep-th/0702086, 2007.

\bibitem{BJMST06b}
\bysame, \emph{Hidden {G}rassmann structure in the {XXZ} model}, Comm. Math.
  Phys. \textbf{272} (2007), 263.

\bibitem{BJMST08app}
\bysame, \emph{Hidden {G}rassmann structure in the {XXZ} model {II}: creation
  operators}, preprint, arXiv:0801.1176, 2008.

\bibitem{BoGo05}
M.~Bortz and F.~G\"ohmann, \emph{Exact thermodynamic limit of short range
  correlation functions of the antiferromagnetic {XXZ} chain at finite
  temperatures}, Eur. Phys. J. B \textbf{46} (2005), 399.

\bibitem{FaMc99}
K.~Fabricius and B.~M. McCoy, \emph{Quantum-classical crossover in the spin-1/2
  {XXZ} chain}, Phys. Rev. B \textbf{59} (1999), 381.

\bibitem{GKS04a}
F.~G\"ohmann, A.~Kl\"umper, and A.~Seel, \emph{Integral representations for
  correlation functions of the {XXZ} chain at finite temperature}, J. Phys. A
  \textbf{37} (2004), 7625.

\bibitem{GKS05}
\bysame, \emph{Integral representation of the density matrix of the {XXZ} chain
  at finite temperature}, J. Phys. A \textbf{38} (2005), 1833.

\bibitem{GoSe05}
F.~G\"ohmann and A.~Seel, \emph{{XX} and {I}sing limits in integral formulae
  for finite temperature correlation functions of the {XXZ} chain}, Theor.
  Math. Phys. \textbf{146} (2006), 119.

\bibitem{Ja12}
Dunham Jackson, \emph{On approximation by trigonometric sums and polynomials},
  Trans. Amer. Math. Soc. \textbf{13} (1912), 491--515.

\bibitem{KSTS03}
G.~Kato, M.~Shiroishi, M.~Takahashi, and K.~Sakai, \emph{Next-nearest-neighbour
  correlation functions of the spin-1/2 {XXZ} chain at the critical region}, J.
  Phys. A \textbf{36} (2003), L337.

\bibitem{KSTS04}
\bysame, \emph{Third-neighbour and other four-point correlation functions of
  spin-1/2 {XXZ} chain}, J. Phys. A \textbf{37} (2004), 5097.

\bibitem{Kluemper92}
A.~Kl\"umper, \emph{Free energy and correlation length of quantum chains
  related to restricted solid-on-solid lattice models}, Ann.\ Physik \textbf{1}
  (1992), 540.

\bibitem{Kluemper93}
\bysame, \emph{Thermodynamics of the anisotropic spin-1/2 {H}eisenberg chain
  and related quantum chains}, Z. Phys. B \textbf{91} (1993), 507.

\bibitem{La50}
Cornelius Lanczos, \emph{An iteration method for the solution of the eigenvalue
  problem of linear differential and integral operators}, J. Res. Nat. Bur.
  Stand. \textbf{45} (1950), 255--282.

\bibitem{TsSh05}
Z.~Tsuboi and M.~Shiroishi, \emph{High temperature expansion of the emptiness
  formation probability for the isotropic {H}eisenberg chain}, J. Phys. A
  \textbf{38} (2005), L363.

\bibitem{Vermaseren00}
J.~A.~M. Vermaseren, \emph{New features of {FORM}}, preprint,
  ar{X}iv:math-ph/0010025, 2000.

\bibitem{WWAF06}
Alexander Wei{\ss}e, Gerhard Wellein, Andreas Alvermann, and Holger Fehske,
  \emph{The kernel polynomial method}, Rev. Mod. Phys. \textbf{78} (2006),
  275--306.

\bibitem{YaYa66a}
C.~N. Yang and C.~P. Yang, \emph{Ground-state energy of a {H}eisenberg-{I}sing
  lattice}, Phys. Rep. \textbf{147} (1966), 303.

\end{thebibliography}

\providecommand{\bysame}{\leavevmode\hbox to3em{\hrulefill}\thinspace}
\providecommand{\MR}{\relax\ifhmode\unskip\space\fi MR }
\providecommand{\MRhref}[2]{%
  \href{http://www.ams.org/mathscinet-getitem?mr=#1}{#2}
}
\providecommand{\href}[2]{#2}

\end{document}